\documentclass[aps,prb,twocolumn,floats,showpacs]{revtex4}
\usepackage{epsfig}
\usepackage{bm}
\usepackage{latexsym}

\begin{document}

\newcommand{\hide}[1]{}
\newcommand{\tbox}[1]{\mbox{\tiny #1}}
\newcommand{\half}{\mbox{\small $\frac{1}{2}$}}
\newcommand{\sinc}{\mbox{sinc}}
\newcommand{\const}{\mbox{const}}
\newcommand{\trc}{\mbox{trace}}
\newcommand{\intt}{\int\!\!\!\!\int }
\newcommand{\ointt}{\int\!\!\!\!\int\!\!\!\!\!\circ\ }
\newcommand{\eexp}{\mbox{e}^}
\newcommand{\bra}{\left\langle}
\newcommand{\ket}{\right\rangle}
\newcommand{\EPS} {\mbox{\LARGE $\epsilon$}}
\newcommand{\ar}{\mathsf r}
\newcommand{\im}{\mbox{Im}}
\newcommand{\re}{\mbox{Re}}
\newcommand{\bmsf}[1]{\bm{\mathsf{#1}}}
\newcommand{\new}[1]{{\bf #1}}


\title{Resonance width distribution for high-dimensional random media }
\author{Matthias Weiss$^{1}$, J. A. M\'endez-Berm\'udez$^{2,3}$, and Tsampikos Kottos$^{2,4}$}
\affiliation{$^1$Cellular Biophysics Group (BIOMS), German Cancer Research
    Center, Im Neuenheimer Feld 580, D-69121 Heidelberg, Germany}
\affiliation{$^2$Max-Planck-Institut f\"ur Dynamik und Selbstorganisation, Bunsenstrasse 10,
D-37073 G\"ottingen, Germany}
\affiliation{$^3$Department of Physics, Ben-Gurion University, Beer-Sheva 84105, Israel}
\affiliation{$^4$Department of Physics, Wesleyan University, Middletown, CT
06459-0155, USA}

\date{\today}

\begin{abstract}
We study the distribution of resonance widths ${\cal P}(\Gamma)$ for 
three-dimensional (3D) random scattering media and analyze how it changes 
as a function of the randomness strength. We are able to identify in 
${\cal P}(\Gamma)$ the system-inherent fingerprints of the metallic,
localized, and critical regimes. Based on the properties of resonance
widths, we also suggest a new criterion for determining and analyzing
the metal-insulator transition. Our theoretical predictions are verified
numerically for the prototypical 3D tight-binding Anderson model.
\end{abstract}

\maketitle

\section{Introduction}
\label{sec:introduction}

Quantum mechanical scattering in systems with complex internal dynamics
has been a subject of intensive research activity for a number of years.
The interest was motivated by various areas of physics, ranging from nuclear
\cite{MW69}, atomic\cite{atomic} and molecular \cite{molecular} physics,
to mesoscopics \cite{B97}, quantum chaos \cite{S89,S99}, and classical
wave scattering \cite{DS90}. Recently, the interest in this subject was
renewed due to technological developments in quantum optics associated
with the construction of new type of lasers \cite{WAL95,NS97} and the
experimental investigation of atoms in optical lattices \cite{Raizen}.

The most fundamental object which characterizes the process of quantum
scattering is the scattering matrix $S$, where $S$ relates the amplitudes
of waves that enter and leave a scattering region. Of great interest are
the statistical properties of the poles of the $S$ matrix. They determine
the conductance fluctuations of a quantum dot in the Coulomb blockade regime
\cite{ABG02} or the current relaxation \cite{AKL91}. The poles of the $S$
matrix are related to resonance states occurring at complex energies ${\cal
E}_n = E_n - \frac i2 \Gamma_n$, where $E_n$ is the position and $\Gamma_n$
the width of the resonance. Resonances correspond to ``eigenstates" of the
open system that decay in time due to the coupling to the ``outside world".

For chaotic systems Random Matrix Theory (RMT) is applicable and the
distributions of resonance widths ${\cal P}(\Gamma)$ is known. A review can
be found in Ref. \onlinecite{FS97} (see also Ref. \onlinecite{KS00}).
As the disorder increases, the system becomes diffusive and the deviations
from RMT increase drastically. For low dimensional random systems
in the metallic regime the distribution of resonances ${\cal P} (\Gamma)$
was found recently \cite{BGS91,OKG02,OKG03}. For the strongly disordered
limit, where localization dominates, ${\cal P} (\Gamma)$ was investigated
by various groups \cite{TF00,SJNS00,PROT04} as well. At the same time an
attempt to understand systems at critical conditions was undertaken in Refs.
\onlinecite{SOKG00} and \onlinecite{KW02}. The latter deals with random
systems of higher dimensions, the most prominent of which is the
three-dimensional (3D) Anderson model. It undergoes a Metal-Insulator
Transition (MIT) with increasing strength of disorder \cite{A58}.

In this paper we extend our previous analysis on the 3D Anderson model \cite{KW02}
and study the distribution of resonance widths as we change the disorder strength.
Based on the analysis of ${\cal P}(\Gamma)$ we propose a new method to locate
the MIT. The paper is organized as follows: In Section \ref{sec:basics} the 3D
Anderson model and the scattering formalism are
introduced. In Sec. \ref{sec:results} we discuss the consequences of
localization in the distribution of resonance widths in the diffusive and
localized regimes as well as at the MIT and show the numerical results
supporting our arguments. In Sec. \ref{sec:scale} we investigate a new method for determining
and analyzing the emergence of the MIT and propose a scaling theory near
the critical point. Finally, our conclusions are given in Sec.
\ref{sec:conclusions}.

\section{The 3D Anderson model and the scattering setup}
\label{sec:basics}

The Anderson model with diagonal disorder on a 3D cubic lattice
is described by the tight-binding Hamiltonian (TBH)
\begin{equation}
\label {tbh}
H_0=\sum_{\bf n} |{\bf n}\rangle W_{\bf n}\langle {\bf n}| + \sum_{{\bf (n,m)}}
|{\bf n}\rangle \langle {\bf m}| \ ,
\end{equation}
where ${\bf n}\equiv (n_x,n_y,n_z)$ labels all the $N=L^3$ sites of the cubic
lattice, while the second sum is taken over all nearest-neighbor pairs
${\bf (n,m)}$ on the lattice. The on-site potential $W_{\bf n}$ for
$1\leq n_x,n_y,n_z\leq L$ is independently and identically distributed with
probability ${\cal P}(W_{\bf n})$. We use three different distributions for
the random potential: (a) a box distribution, i.e., the $W_{\bf n}$ are
uniformly distributed on the interval $[-W/2, W/2]$; (b) a Gaussian distribution
with zero mean and variance $W^2/12$; and (c) a Cauchy distribution
${\cal P}(W_{\bf n})=W/\pi (W_{\bf n}^2+W^2)$. For the system defined by
Eq.~(\ref{tbh}) the MIT for $E\simeq 0$ occurs for $W=W_c$ with (a)
$W_c\simeq 16.5$, (b) $W_c\simeq 21.3$, and (c) $W_c\simeq 4.26$ (see
Ref.~\onlinecite{RMS01}). Then, for $W<W_c$ ($W>W_c$) the system is in the
metallic (insulating) regime.

\begin{figure}
\begin{center}
    \epsfxsize=6.4cm
    \leavevmode
    \epsffile{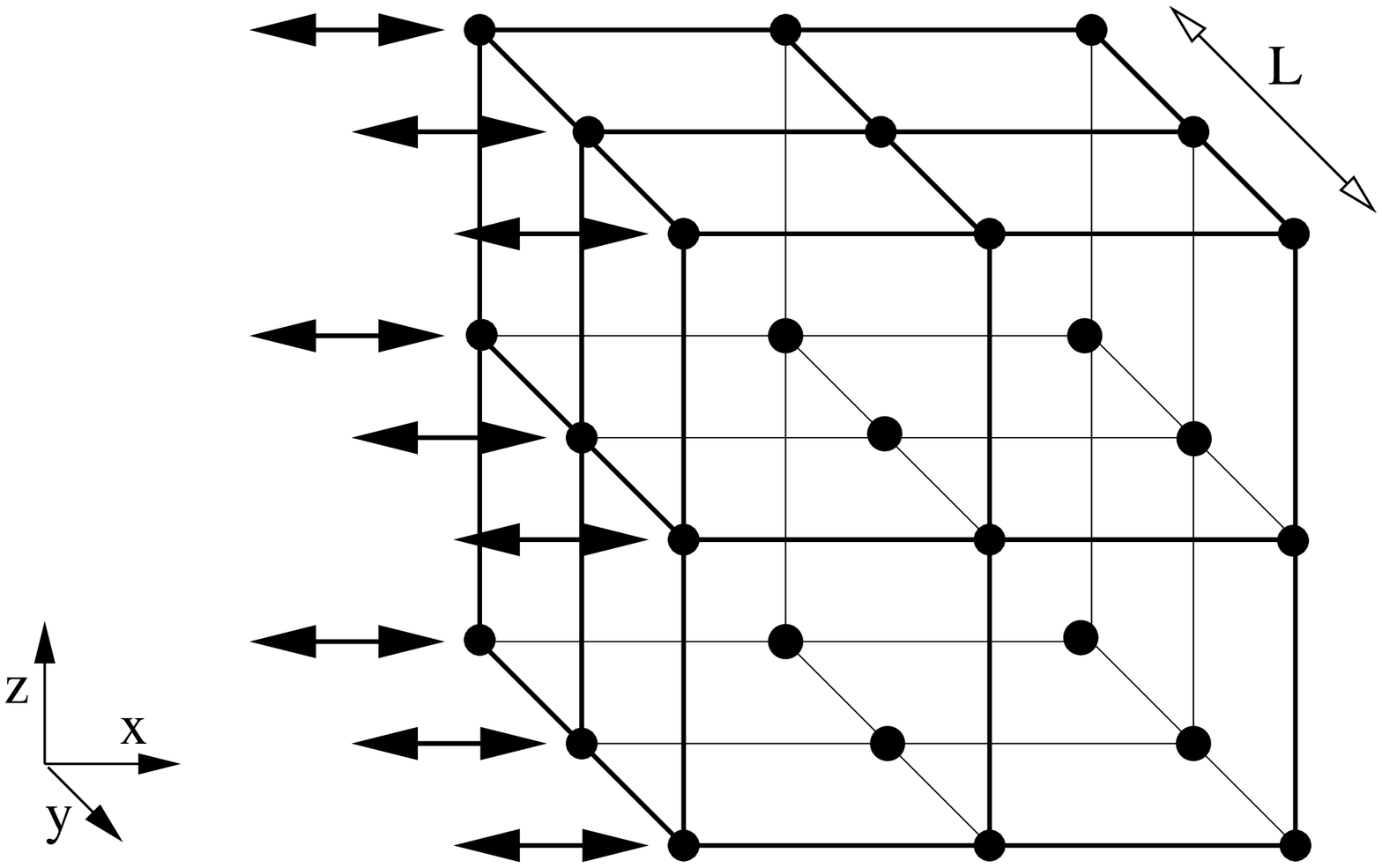}
\caption{Scattering setup. The sample is a cubic lattice of linear length $L$.
To each of the $M=L^2$ sites of the layer $n_x=1$ semi-infinite single mode leads
are attached.}
\label{fig:setup}
\end{center}
\end{figure}

We turn the isolated system to a scattering one by attaching $M=L^2$ semi-infinite single
mode leads to each site of the layer $n_x=1$, as depicted in Fig. \ref{fig:setup}.
Each lead is described by a one-dimensional semi-infinite TBH
\begin{equation}
\label{leads}
H_M=\sum^{-\infty}_{n=1} (|n><n+1| + |n+1><n|) \ .
\end{equation}

Using standard methods \cite{MW69} one can write the scattering matrix in the form
\cite{SOKG00,KW02}
\begin{equation}
\label{smatrix}
S(E) = {\bf 1}-2i \sin (k)\, {\cal W}^{\,T} (E-{\cal H}_{\rm eff})^{-1} {\cal W} \ ,
\end{equation}
where ${\bf 1}$ is the $M\times M$ unit matrix, $k=\arccos(E/2)$ is the wave
vector supported in the leads, and ${\cal H}_{\rm eff}$ is an effective
non-hermitian Hamiltonian given by
\begin{equation}
\label{Heff}
{\mathcal{H}}_{\rm eff}=H_0- e^{ik} {\cal W}{\cal W}^{\,T} \ .
\end{equation}
Here, ${\cal W}$ is a $N\times M$ matrix that specifies at which site of the sample we
attach the leads. Its elements are equal to zero or $\sqrt{w}$, with $0<\sqrt{w}\le 1$,
where $w$ is the coupling strength. Below, unless stated
otherwise, we will always consider the case $w=1$.
Moreover, since $\arccos (E/2)$ changes only slightly in the center of the band,
we set $E=0$ and neglect the energy dependence of ${\mathcal{H}}_{\rm eff}$.
The poles of the $S$ matrix are then equal to the complex zeros of
\begin{equation}
\label{poleseq}
\det [{\cal E}-H_{\rm eff}]=0.
\end{equation}
>From Eqs.~(\ref{smatrix}) and (\ref{poleseq}) it is clear that the formation of
resonances is closely related to the dynamics in the scattering region, governed
by $H_0$.

In order to investigate the distributions of resonance widths
we used samples with $L=20$ as a maximum size. For better statistics a
considerable number of different disorder realizations was considered.
In all cases we had at least 10 000 data for statistical processing.

\section{Distribution of resonance widths: Results and discussion}
\label{sec:results}

\subsection {Metallic Regime}
\label{subsec:difres}

When the disorder strength $W$ is smaller than $W_c$, but still large enough so that
the mean free path is smaller than the system size, the system is in the metallic regime.

Recently, a lot of research activity was devoted to the understanding
of the statistical properties of various physical quantities (such as
conductance, local density of
states, current relaxation times) in finite-size random systems in the metallic regime.
The outcome of these studies indicated that the tails of these distribution functions
show large deviations from the universal Random Matrix Theory (RMT) results, expected
to be valid \cite{E83} in the limit of infinite dimensionless conductance
$g=\Gamma_{\rm Th}/ \Delta=D L$. Here, $\Gamma_{\rm Th} \sim D/L^2$ is the typical
inverse time (Thouless time) that an excitation needs to diffuse (with diffusion
coefficient $D$) in order to reach the boundary of a system, with linear size
$L$, and $\Delta\sim 1/L^3$ is the mean level spacing.

The origin of these deviations was found to be related to the existence of
eigenstates which are unusually localized around a center of localization.
These states are precursors of the Anderson localization and were termed {\it
prelocalized} states \cite{M00,FE95,KOG03,UMRS00}. In 3D conductors they have
sharp amplitude peaks on the top of a homogeneous background \cite{M00,FE95}.

We start our analysis by investigating the effects of prelocalized states in the
distribution of resonance widths. It is natural to expect that these states with
localization centers at the bulk of the sample are affected only weakly when opening 
the system at the boundaries. Therefore, prelocalized states decay very slowly to the
continuum leading us to the conclusion that the corresponding resonance widths
(inverse lifetime) $\Gamma$ are smaller than the mean level spacing $\Delta$. Hence,
assuming the validity of standard first order perturbation theory (that can be applied
if the coupling of the sample to the leads is weak, $w\ll 1$) we get
\begin{equation}
\label{pertgamma}
{\Gamma\over 2} = \langle \Psi|{\cal W}^{\dagger}{\cal W}|\Psi\rangle \propto
\sum_{n\in {\rm boundary}}|\Psi(n)|^2 \sim L^2 |\Psi(L)|^2 \ ,
\end{equation}
where $|\Psi (L)|^2$ is the wavefunction intensity of a pre-localized state at
the boundary of the sample. At the same time, the distribution of wavefunction
components at the boundary was found to be \cite{FE95}
\begin{equation}
\label{theta}
{\cal P}(\theta) \sim \exp\left[-C_1\ln^3 \left(\theta\right)\right] \ ,
\end{equation}
with $\theta^{-1}=L\Psi(L)$ and $C_1 \propto g$. Using Eq.~(\ref{theta})
together with Eq.~(\ref{pertgamma}) we obtain
\begin{equation}
\label{difgammaS}
{\cal P}(1/\Gamma) \sim \exp\left[-C_2 \ln^3 (1/\Gamma)\right] \ ,
\end{equation}
where $C_2 \propto g$.

\begin{figure}
\begin{center}
    \epsfxsize=8.4cm
    \leavevmode
    \epsffile{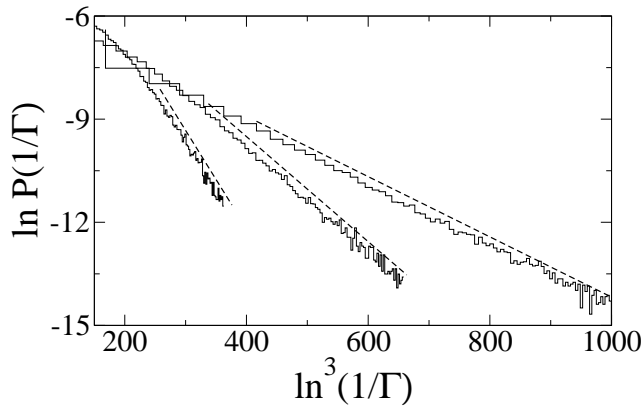}
\caption{The distribution of resonance widths, plotted as ${\cal P}(1/\Gamma)$
vs $\ln^3(1/\Gamma)$, for $\Gamma < \Delta$ in the diffusive regime.
$L=16$ and $W=10$, 12, and 14 (from left to right).
The dashed lines are (shifted) linear fittings to the distributions.}
\label{fig:Fig2}
\end{center}
\end{figure}

As can be seen from Fig.~\ref{fig:Fig2} the prediction of Eq.~(\ref{difgammaS}),
obtained using perturbation theory, holds even for strong coupling. Indeed,
the reported data for the 3D Anderson model in the diffusive regime, plotted
as $\ln {\cal P}(1/\Gamma)$ vs $\ln^3(1/\Gamma)$, shows a linear behaviour.
This comes as a surprise since in Fig.~\ref{fig:Fig2} we have considered
perfect coupling, $w=1$.
Differences in slope correspond to different dimensionless conductances $g$
induced by the varying values of disorder strengths.

Next, we turn to the analysis of ${\cal P}(\Gamma)$ for
$\Gamma \gtrsim \Gamma_{\rm Th}\gg \Delta$.
In order to go on we need to recall that
the inverse of $\Gamma$ represents the quantum lifetime of a particle in a resonant state
escaping into the leads. Moreover, we assume that the particles are uniformly distributed
inside the sample and spread until they reach the boundary where they are absorbed. Then,
we can associate the corresponding lifetimes with the time $t_R\sim 1/ \Gamma_R $ a particle
needs to reach the boundaries, when starting a distance $R$ away. The relative number of
states that require a time $t<t_R$ in order to reach the boundaries (or equivalently the
number of states with $\Gamma >\Gamma_R$) is
\begin{equation}
\label{igam}
{\cal P}_{\rm int}(\Gamma_R)=\int_{\Gamma_R}^{\infty}{\cal P}(\Gamma)d\Gamma
\sim\frac{V(t_R)}{L^3} \ ,
\end{equation}
where $V(t_R)\sim L^3-(L-R)^3$ is the volume populated by all particles with
lifetimes $t<t_R$.

Assuming now diffusive spreading,
\[
R^2=D\cdot t_R \ ,
\]
we get from Eq.~(\ref{igam}) (to leading order with respect to $\Gamma_{\rm Th}/\Gamma$)
\begin{equation}
\label{gamlargedif}
{\cal P}({\widetilde \Gamma}) \sim \left({{\widetilde \Gamma_{\rm Th}}\over {\widetilde \Gamma}}\right)^{3/2} \
=\left({g\over{\widetilde \Gamma}}\right)^{3/2}
\end{equation}
where we refer to the rescaled variable $\widetilde \Gamma = \Gamma/\Delta$.
Equation (\ref{gamlargedif}) is valid as long as the leads are attached
to the boundary of the sample.

Here, it is interesting to point out that a different way of opening the system might
lead to a different power law behavior for ${\cal P}(\Gamma)$. Such a situation can
be realized if instead of opening the system at the boundaries we attach one lead
somewhere in the sample. In such a case we have
\[
{\cal P}_{\rm int}(\Gamma_R) \sim \frac{V(t_R)}{L^3} \approx \frac{R^3}{L^3} =
\frac{(D \cdot t_R)^{3/2}}{L^3} \sim \left( \frac{\Gamma_{\rm Th}}{\Gamma_R}
\right)^{3/2} \ ,
\]
leading to
\begin{equation}
\label{finalg2}
{\cal P}(\Gamma) \sim \Gamma^{-5/2} \ ,
\end{equation}
where we used $V(t_R)\sim R^3$. The above results are correct for any number
of leads $M$ such that the ratio $M/L^3$ scales as $1/L^3$.

If, on the other hand, we attach a single lead to the boundary of the sample we obtain
\begin{equation}
\label{finalg1a}
{\cal P}(\Gamma) \sim \Gamma^{-2} \ .
\end{equation}
This behavior is due to the fact that the decay takes place at the surface, leading 
to a situation similar to that of a 2D system \cite{OKG03}.

\begin{figure}
\begin{center}
    \epsfxsize=8.4cm
    \leavevmode
    \epsffile{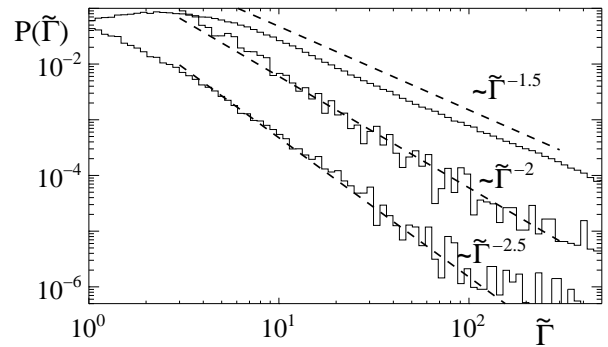}
\caption{The resonance width distribution ${\cal P}({\widetilde \Gamma})$
for $L=16$ and various scattering configurations (from up to down): $M=L^2$
leads attached to the boundary, one lead attached to the boundary, and one
lead attached to a site in the bulk of the sample. The dashed lines are the 
corresponding theoretical predictions for $\Gamma \gtrsim \Gamma_{\rm Th}$ 
given by Eqs.~(\ref{gamlargedif}-\ref{finalg1a}), see main text.}
\label{fig:metfig1}
\end{center}
\end{figure}

In Fig. \ref{fig:metfig1} we present numerical data for the 3D Anderson
model in the metallic regime. We verify the validity of the
theoretical predictions given by Eqs. (\ref{gamlargedif}-\ref{finalg1a})
by using various configurations: $M=L^2$ leads attached to the boundary,
one lead attached to the boundary, and one lead attached to a site in
the bulk of the sample, respectively. We observe a good agreement with
the expected behavior in all cases.

\subsection {Localized Regime}
\label{subsec:locres}

When the disorder strength $W$ is larger than $W_c$ the system is in
the localized regime. In this regime the eigenfunctions are exponentially
localized in space and, as a consequence, transmission is inhibited and
the system behaves as an insulator.

Various groups \cite{TF00,SJNS00,PROT04} had investigated the resonance width
distribution of low dimensional random media in the localized regime during
the last years. In the region of exponentially narrow resonances $\Gamma<\Gamma_0
=\exp(-2L/l_{\infty})$ the distribution was found to be log-normal, i.e.,
\begin{equation}
\label{locres2}
{\cal P}(\widetilde \Gamma)\sim \exp\left[-\left(4 {L\over l_{\infty}}\right)^{-1}
\ln^2(\widetilde \Gamma)\right] , \quad  \Gamma< \Gamma_0 \ .
\end{equation}
This result is analogous to the conductance distribution of localized systems.
Equation (\ref{locres2}) essentially relies on two assumptions: first, that
eigenfunction components are randomly distributed with no long-range correlations;
and second, that they are exponentially localized with a normal distribution of
localization lengths.

\begin{figure}
\begin{center}
    \epsfxsize=8.4cm
    \leavevmode
    \epsffile{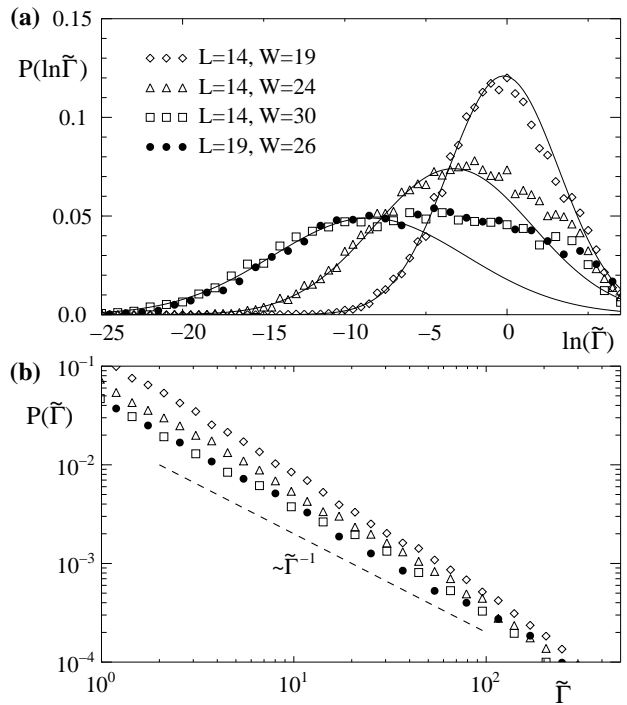}
\caption{\label{loc}${\cal P}(\widetilde{\Gamma})$ in the localized regime for
various combinations of $W$ and $L$ in the range $\widetilde{\Gamma}\le 1$.
The log-normal decay is highlighted by Gaussian fits (full curves) whose maximum
decreases with increasing strength of disorder and also shifts towards smaller
values of $\widetilde{\Gamma}$.
When keeping the ratio $l_\infty/L\approx0.136$ fixed, similar distributions
were obtained for different combinations of $L$ and $W$ (filled circles and open
squares).
(b) For $\widetilde{\Gamma}\ge 1$ the power-law decay
${\cal P}(\widetilde{\Gamma})\sim 1/\widetilde{\Gamma}$ is observed (dashed line)
which  becomes more robust for increasing strength of disorder.}
\label{fig:locfig1}
\end{center}
\end{figure}

It is reasonable to assume that the same arguments leading to Eq.~(\ref{locres2})
applies as well for high-dimensional random media like the 3D Anderson model in the
localized regime. Indeed, our numerical results reported in Fig.~\ref{fig:locfig1}(a)
show good agreement with the theoretical expectation (\ref{locres2})
which supports our assumption.

On the other hand, we found [see Fig.~\ref{fig:locfig1}(b)] that the long tails
of  the distribution behave as
\begin{equation}
\label{locres1}
{\cal P}({\widetilde \Gamma})\sim \left({l_{\infty}\over L}\right)
{1\over {\widetilde \Gamma}},\quad  \Gamma_0< \Gamma \ll 1/L \ .
\end{equation}
Equation (\ref{locres1}) can be easily understood when employing Eq.~(\ref{igam}).
The new ingredient is that wavefunctions are exponentially localized:
$|\Psi(r)|\sim l_{\infty}^{-3/2} \exp(-r/l_{\infty})$. Using simple perturbation
arguments \cite{TF00} [see Eq.~(\ref{pertgamma})], i.e. $\Gamma \sim |\Psi(r)|^2$, we obtain
\[
R^3 \sim l_{\infty}^3 \ln^3(l_{\infty}^3\Gamma) \ .
\]
By inserting this into Eq.~(\ref{igam}) we get Eq.~(\ref{locres1}), to leading order
with respect to $l_{\infty}/L$.

The region of large $\Gamma$ values is essentially determined by the
coupling to the continuum,
so it should be model-dependent. Nevertheless, it is reasonable to assume that the
number of resonances involved is constant, of order $l_{\infty}$, and therefore the
extreme tail of the distribution should subside at large $L$ at the rate
$\sim l_{\infty}/L$.

Let us finally note that in the thermodynamic limit $L\rightarrow \infty$ the
probability of finding an eigenstate at any finite distance from the boundary
is equal to zero. Thus the distribution of the resonance widths in this case
approaches a delta function centered at zero.

\subsection {Criticality}
\label{subsec:critres}

The MIT, where $W=W_c$, is characterized by several {\it critical} properties:
the level statistics acquires a scale-independent form \cite{M00,SSSLS93,AS86,CKL96}
while the eigenfunctions show strong fluctuations on all length scales and
obey multifractal distributions \cite{M00,FE95,W80,SG91,EM00}.

In Ref. \onlinecite{KW02} it was found that ${\cal P}(\widetilde \Gamma)$ follows
a new universal distribution, i.e., independent of the microscopic details of the
random potential and number of attached leads.
Specifically, it decays asymptotically with the power
\begin{equation}
\label{MITgam}
{\cal P}({\widetilde \Gamma})\sim g_{\rm c}^{1/3}  {\widetilde \Gamma}^{-(1+1/3)} \ ,
\end{equation}
which is different from those found for chaotic, metallic, or localized systems
(see Ref. \onlinecite{FS97}, Eq.~(\ref{gamlargedif}), and Eq.~(\ref{locres1}),
respectively).

One can relate the power-law decay (\ref{MITgam}) to the anomalous diffusion at the
MIT. Indeed, at the MIT the conductance of a 3D disordered sample has a
finite value $g_{\rm c}\sim 1$. Approaching the MIT from the metallic regime one has
$g\sim E_T/ \Delta$, where $E_T=D/R^2$ is the Thouless energy, $D$ is the diffusion
coefficient, and $\Delta \sim 1/R^3$ is the mean level spacing of a
sample with linear size $R$. This yields $D\sim g_{\rm c}/R$ at $W_c$.
Taking into account that $D=R^2/t_R$, we get for the spreading of an excitation at
the MIT
\[
R^3 \sim g_{\rm c} \cdot t_R \ .
\]
Then, straightforward application of Eq.~(\ref{igam}) leads to Eq.~(\ref{MITgam}).

\begin{figure}
\begin{center}
    \epsfxsize=8.4cm
    \leavevmode
    \epsffile{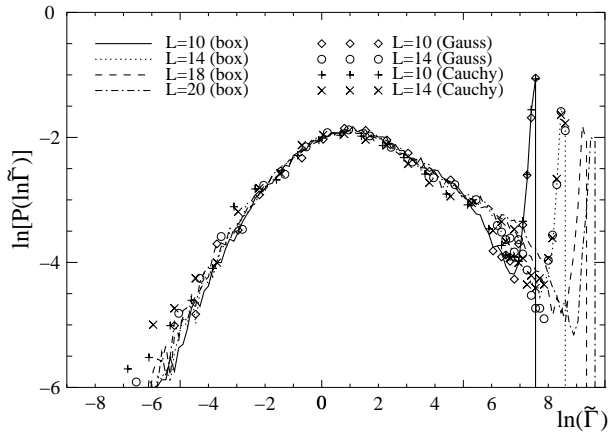}
\caption{Universal behavior of ${\cal P}({\widetilde \Gamma})$ at the MIT [reported
here as ${\cal P}(\ln({\widetilde \Gamma}))$] for various sample sizes $L$ and potential
distributions.}
\label{fig:crit1}
\end{center}
\end{figure}

\begin{figure}
\begin{center}
    \epsfxsize=8.4cm
    \leavevmode
    \epsffile{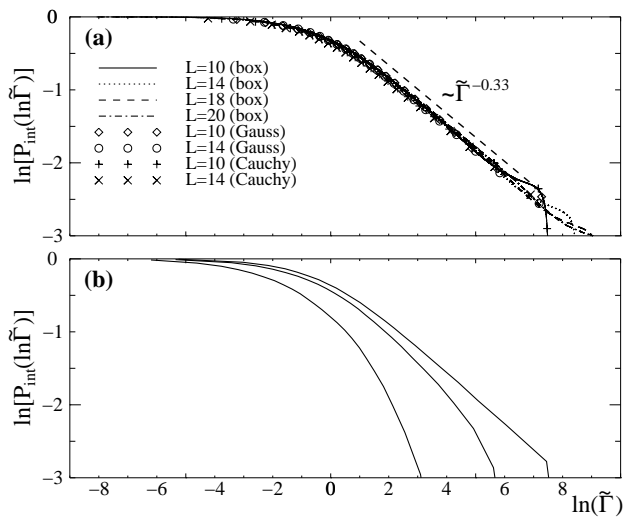}
\caption{(a) The integrated distribution ${\cal P}_{\rm int}({\widetilde \Gamma})$
in the case of perfect coupling, $w=1$, for various sample sizes $L$ and potential
distributions. The dashed line is the theoretical prediction
${\cal P}_{\rm int}({\widetilde \Gamma}) \sim {\widetilde \Gamma}^{-0.333}$, see
Eq. (\ref{MITgam}).
(b) ${\cal P}_{\rm int}({\widetilde \Gamma})$ for a box distribution, $L=10$, and
different coupling strengths: $w=0.001$, 0.01, and 0.5 from left to right.}
\label{fig:crit2}
\end{center}
\end{figure}

In Figs.~\ref{fig:crit1} and \ref{fig:crit2} we report some numerical results for
the 3D Anderson model at the MIT.
Figure \ref{fig:crit1} shows the distribution of the logarithm of the rescaled
resonance widths ${\cal P}(\ln({\widetilde \Gamma}))$ for the three
different distributions ${\cal P}(W_{\bf n})$ of the random potential and for
various sample sizes $L$. The body of the distribution function in all cases
coincides and does not change its shape or width. Of course, the far tail of
this universal distribution develops better with increasing $L$.
The sharp peak appearing at the right is an artifact of our choice to neglect
the energy dependence of ${\cal H}_{\rm eff}$. We thus confirm that at the
MIT the distribution of rescaled resonances is indeed scale-invariant independent
of the microscopic details of the potential.

>From Fig. \ref{fig:crit2}(a) an inverse power law ${\cal P}_{ \rm int}({\widetilde
\Gamma}) \sim {\widetilde \Gamma}^{ -\alpha}$ is evident. The best fit to the numerical
data yields $\alpha= 0.333\pm 0.005$ in accordance with Eq.~(\ref{MITgam}). Here,
the case of perfect coupling ($w=1$) has been considered. Different coupling strengths
are going to affect this behavior as can be seen from Fig. \ref{fig:crit2}(b).

\section{ A scaling Theory for the Resonances Widths}
\label{sec:scale}

In the original proposal of the scaling theory of localization, the conductance $g$
is the relevant parameter \cite{A58,AALR79}. A manifestation of this statement is seen
in Eqs.~(\ref{difgammaS},\ref{gamlargedif},\ref{locres2},\ref{locres1},\ref{MITgam}) where
${\cal P}_{\rm int}^{0}\equiv {\cal P}_{\rm int}({\widetilde \Gamma}_0)$ is proportional
to the conductance $g$. It is therefore natural to expect that ${\cal P}_{\rm int}^{0}$
will follow a scaling behavior for finite $L$ (and for some ${\widetilde \Gamma}_0\sim 1$)
that is similar to the one obeyed by the conductance $g$.
The following scaling hypothesis was therefore postulated in Ref. \onlinecite{KW02}:
\begin{equation}
\label{scale}
{\cal P}_{\rm int}^{0}(W,L)=f(L/l_{\infty}(W)) \ .
\end{equation}
In the insulating phase
($W>W_c$) the conductance of a sample with length $L$ behaves as
$g(L)\sim \exp(-L/l_\infty)$ due to the exponential localization of the eigenstates,
and therefore we have $g(L_1) < g(L_2)$ for $L_1>L_2$. Based on Eq.~(\ref{MITgam})
we expect the same behavior for ${\cal P}_{\rm int}^{0}$; i.e., for every finite $L_1>L_2$
we must have ${\cal P}_{\rm int}^{0}(W,L_1)<{\cal P}_{\rm int}^{0}(W,L_2)$. On the other
hand, in the metallic regime ($W<W_c$) we have that $g(L)=D L$ and therefore we expect
from Eq.~(\ref{MITgam}) ${\cal P}_{\rm int}^{0}(W,L_1)>{\cal P}_{\rm int}^{0}(W,L_2)$.
Thus, the critical point is the one at which the size effect changes its sign, or in other
words, the point where all curves ${\cal P}_{\rm int}^{0}(W,L)$ for various $L$ cross.
One can reformulate the last statement by saying that in the thermodynamic limit
$L\rightarrow \infty$ at $W=W_c$ the number of resonances with width larger
than the mean level spacing goes to a constant.

\begin{figure}
\begin{center}
    \epsfxsize=5.4cm
    \leavevmode
    \epsffile{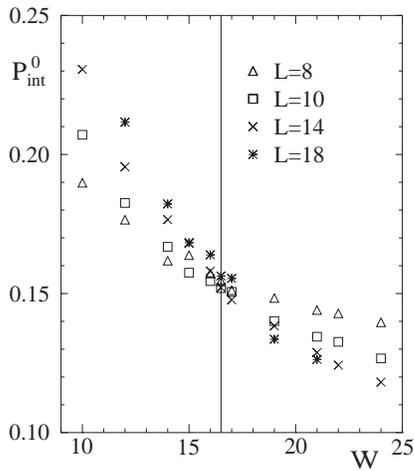}
\caption{${\cal P}_{\rm int}^0(W,L)$ as a function of $W$ for different
system sizes $L$ provides a means to determine the critical point $W_c$ of the MIT
(vertical line at $W=16.5$).}
\label{fig:fig3a}
\end{center}
\end{figure}

\begin{figure}
\begin{center}
    \epsfxsize=5.4cm
    \leavevmode
    \epsffile{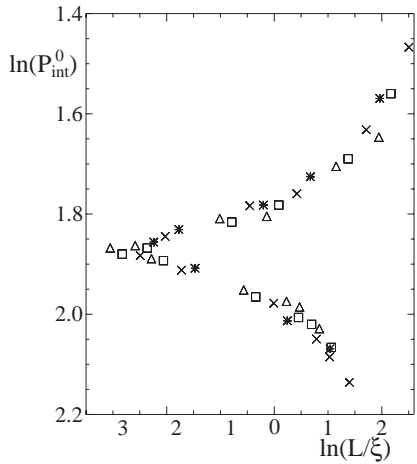}
\caption{The one-parameter scaling of ${\cal P}_{\rm int}^0(W,L)$
[Eq.~(\ref{scale})] is confirmed for various system sizes $L$ and
disorder strengths $W$ using the box distribution.}
\label{fig:fig3b}
\end{center}
\end{figure}

In Fig. \ref{fig:fig3a}, we show the evolution of ${\cal P}_{\rm int}^{0}(W)$ for
different values of $L$ using the box distribution. From this analysis the
critical disorder strength $W=W_c=16.5\pm 0.5$ was determined in agreement with other
calculations\cite{RMS01}. A further verification of the scaling hypothesis
(\ref{scale}) is shown in Fig.~\ref{fig:fig3b} where the same data are reported as a
function of the scaling ratio $L/l_{\infty}$. Note that all points collapse on two
separate branches for $W<W_c$ and $W>W_c$.

\section{Conclusions}
\label{sec:conclusions}

We have studied the properties of the resonance width distribution in 
various regimes for the 3D Anderson model.

In the metallic regime we obtained the forms of ${\cal P}(\Gamma)$ for small and large
$\Gamma$ and show that they are determined by the underlying diffusive dynamics and by
the existence of prelocalized states. For the localized regime we also explored the
limits of small and large $\Gamma$. In the first limit we found that ${\cal P}(\Gamma)$
shows a log-normal behavior while in the latter the distribution is power-law like.
At the MIT we show that ${\cal P}({\widetilde \Gamma})$, with
${\widetilde \Gamma}=\Gamma/\Delta$, has a {\it universal} form, i.e.,
independent of the microscopic details of the random potential and number
of attached leads. Specifically, it decays asymptotically with a power
which is different from those found in the diffusive and localized regimes.
In addition, based on resonance widths, we suggested a new method for determining 
and analyzing the emergence of the MIT and propose a scaling theory near the critical
point.  

\vspace*{0.5cm}
\section{Acknowledgments}

T. K. acknowledges many useful comments and discussions with Prof. B. Shapiro.
J.A.M.-B. and T.K. acknowledge support by a grant from the GIF, the German-Israeli
Foundation for Scientific Research and Development. M.W. acknowledges support from
the BIOMS initiative in Heidelberg.


\end{document}